\documentstyle[epsfig]{mn}
\newcommand{\figurepath}{./}
\newcommand{\bx}{_{\mbox{\tiny box}}}
\newcommand{\tl}{_{\mbox{\tiny tot}}}
\newcommand{\br}{_{\mbox{\tiny elb}}}

\newcommand{\bn}{_{\mbox{\tiny bin}}}
\newcommand{\cl}{_{\mbox{\tiny clu}}}
\newcommand{\oc}{_{\mbox{\tiny occ}}}
\newcommand{\pk}{_{\mbox{\tiny peak}}}
\newcommand{\mn}{_{\mbox{\tiny min}}}
\newcommand{\mx}{_{\mbox{\tiny max}}}

\begin{document}

\title{The structure of young star clusters}

\author[Gladwin et al.]
  {P. P. ~Gladwin\thanks{e-mail: philip.gladwin@astro.cf.ac.uk},
  S. ~Kitsionas\thanks{e-mail: spyridon.kitsionas@astro.cf.ac.uk},
  H. M. J. ~Boffin\thanks{e-mail: henri.boffin@astro.cf.ac.uk}
  and A. P.~Whitworth\thanks{e-mail: ant.whitworth@astro.cf.ac.uk}\\
  Department of Physics and Astronomy, University of Wales, Cardiff, CF2 3YB, Wales, UK.
  }

\maketitle
\begin{abstract}
In this paper we analyse and compare the clustering of young stars in 
Chamaeleon I and Taurus. We compute the mean 
surface-density of companion stars $\bar{N}$ as a function of 
angular displacement $\theta$ from each star. We then fit $\bar{N}(\theta)$ 
with two {\it simultaneous} power laws, i.e. $\bar{N}(\theta) 
\simeq K\bn \theta^{-\beta\bn} + K\cl \theta^{-\beta\cl}$. For Chamaeleon I, we obtain $\beta\bn = 1.97 \pm 0.07$ and $\beta\cl = 0.28 \pm 0.06$, with the elbow at $\theta\br \simeq 0.011 ^{\circ} \pm 0.004 ^{\circ}$. For Taurus, we obtain $\beta\bn = 2.02 \pm 0.04$ and $\beta\cl = 0.87 \pm 0.01$, with the elbow at $\theta\br \simeq 0.013^{\circ} \pm 0.003 ^{\circ}$.  For both star clusters the observational data make quite large ($\sim 5 \sigma$) systematic excursions from the best fitting curve in the binary regime ($\theta < \theta\br$). These excursions are visible also in the data used by Larson and Simon, and may be attributable to evolutionary effects of the types discussed recently by Nakajima {\it et al.} and Bate {\it et al}. In the clustering regime ($\theta > \theta\br$) the data conform to the best fitting curve very well, but the $\beta\cl$ values we obtain differ significantly from those obtained by other workers. These differences are due partly to the use of different samples, and partly to different methods of analysis.

We also calculate the box-dimensions for the two star clusters: for Chamaeleon I we obtain $D\bx \simeq 1.51 \pm 0.12$, and for Taurus $D\bx \simeq 1.39 \pm 0.01$. However, the limited dynamic range makes these estimates simply descriptors of the large-scale clustering, and not admissible evidence for fractality.

We propose two algorithms for objectively generating maps of constant stellar surface-density in young star clusters. Such maps are useful for comparison with molecular-line and dust-continuum maps of star-forming clouds, and with the results of numerical simulations of star formation. They are also useful because they retain information which is suppressed in the evaluation of $\bar{N}(\theta)$. Algorithm I ({\it SCATTER}) uses a universal smoothing length, and therefore has a restricted dynamic range, but it is implicitly normalized. Algorithm II ({\it GATHER}) uses a local smoothing length, which gives it much greater dynamic range, but it has to be normalized explicitly. Both algorithms appear to capture well the features which the human eye sees. We are exploring ways of analyzing such maps to discriminate between fractal structure and single-level clustering, and to determine the degree of central condensation in small-$N$ clusters. 
\end{abstract}


\section{Introduction}

\subsection{Motivation}

As the number of young stars detected in nearby star-formation regions (SFRs) has increased over the last decade, it has become possible to study the clustering properties in these regions \cite{gomez2,larson,simon2,nakaj,bate,brand}. The manner in which young stars cluster places important constraints on theories of star formation. Numerical simulations of star formation are now approaching a level of sophistication which enables them to make detailed predictions about the density- and velocity-fields in star-forming gas and the distributions of the resulting stars (e.g. Turner et al. (1995), Whitworth et al. (1995), Bonnell et al. (1997), Klessen (1997), Bonnell, Bate \& Zinnecker (1998), Boffin et al. (1998)). Since there is a great diversity of star-forming molecular clouds and young star clusters, useful comparisons between observation and simulation can probably only be made using global statistical descriptors, and this paper seeks to develop such descriptors.

\subsection{Background} 

Speckle interferometry, lunar occulation and spectroscopic surveys have enabled astronomers to estimate the binary fraction in young stellar populations. In general, the proportion of binaries is at least as high as for mature stars in the field, and in Taurus it appears to be even higher. At the same time, near infrared imaging and X-ray surveys of SFRs have revealed many additional young stars which were not detected by the earlier emission-line searches.

Extending earlier work by Gomez et al. (1993), Larson (1995) has computed the mean surface-density of companions per star, $\bar{N}(\theta)$, as a function of angular separation, $\theta$, in Taurus. Since two distinct power laws of the form $\bar{N}(\theta) \sim \theta^{-\beta}$ are necessary to fit his data, Larson concludes that there are two different clustering regimes. The smaller separations define the regime in which binary and higher-order multiple systems are found; in this regime he finds $\beta\bn \sim$ 2.15. The larger separations define the regime in which stars appear to be hierarchically clustered, and here he finds $\beta\cl \sim$ 0.62. Larson notes that the break between the two regimes (hereafter referred to as the elbow) occurs at $s\br \sim$ 0.04 pc, and suggests that this is the Jeans length in the cloud cores from which the stars have formed.

Simon (1997) has repeated this analysis for Taurus, and extended it to the Orion-Trapezium and $\rho$ Ophiuchus SFRs. He finds values of $\beta\bn$ and $\beta\cl$ similar to those found by Larson (1995). However,  the elbow shifts to smaller separations going from Taurus ($s\br \simeq$ 0.06 pc), through $\rho$ Ophiuchus (0.024 pc) to Orion-Trapezium (0.002 pc). Simon argues that it is difficult to explain this shift simply on the basis of a 30-fold variation in the Jeans length. He points out that in most cases the elbow is determined largely by the background surface-density of the overall cluster; this is reflected in the fact that $s\br$ correlates with the mean separation between stars. Essentially the same conclusion was reached by Kitsionas, Gladwin \& Whitworth (1998).

Nakajima et al. \shortcite{nakaj} have calculated $\bar{N}(\theta)$ for the Orion, Ophiuchus, Chamaeleon, Vela and Lupus SFRs. For SFRs where two power laws are required to fit the data, they obtain $\beta\bn \simeq 2.0$ and $0.1 \la \beta\cl \la 0.8$. They show that the elbow occurs at a separation roughly equal to one tenth of the mean nearest-neighbour distance. They also point out that structure which is present when the stars first form will subsequently be erased by random migration due to the velocity dispersion, which is typically $\sim$ 1 km s$^{-1}$. Thus, as unbound clusters disperse, the elbow should shift to large separations. In addition, some SFRs show evidence that star formation has gone on for several Myrs; and so there may be a young, highly clustered population mixed in with an older, less highly clustered population. 

Bate, Clarke \& McCaughrean \shortcite{bate} have recently undertaken a comprehensive evaluation of the factors influencing $\bar{N}(\theta)$ and its interpretation. They show that $\bar{N}(\theta)$ can be strongly affected by boundary effects. They point out that even if $\bar{N}(\theta)$ can be fit with a non-integer power-law, this does not necessarily imply fractality. They illustrate this by considering non-fractal distributions for which $\bar{N}(\theta)$ can nonetheless be fit with a non-integer power-law. In particular, a simple distribution of single-level clusters in a lower-density background can produce an $\bar{N}(\theta)$ which is well fit by a non-integer power-law. They also demonstrate that evolutionary effects -- dispersal of unbound clusters due to their intrinsic velocity dispersion, evaporation of stars from small-$N$ bound clusters, tidal dissolution of wide binaries -- should all act to suppress $\bar{N}(\theta)$ in the vicinity of the elbow.

Whitworth, Boffin \& Francis (1998) have argued that binary formation via dynamical fragmentation only occurs when the gas becomes thermally coupled to the dust. This leads to a critical density below which binaries are unlikely to form via dynamical fragmentation, and hence to a maximum initial binary separation $\sim 0.05$ pc. They suggest that this may contribute to the break in $\bar{N}(\theta)$ at around this separation in sparse star clusters like Taurus, where the overall density is too low for chance projections to be important.

\subsection{Plan}

In this paper  we concentrate on the Chamaeleon I SFR, a region of low mass star formation at a distance of $\sim$ 140 pc \cite{schwartz}. In Section \ref{sec:surf_den} we evaluate $\bar{N}(\theta)$ and the box-dimension of the star field, and in Section 3 we review briefly the procedure used by Gomez et al. (1993) to generate surface-density maps of young star clusters. In Sections 4 and 5 we describe and evaluate two new algorithms that create contour maps of stellar surface-density. In Section 6 we compare our results for Chamaeleon I with Taurus (also at a distance of $\sim$ 140 pc (Elias 1978)), and in Section 7 we summarize our conclusions.

\begin{table*}
\begin{minipage}{105mm}
\caption{Summary of references used to compile the Taurus catalogue. }
\label{tab:refer}
\begin{tabular}{@{}lclc@{}} 
Reference                 & Number of stars & Survey Type & Notes   \\ \hline
Herbig and Bell (1988)    & 118 & Slit spectroscophy & (1)          \\
Walter et al. (1988)      & 14  & X-ray survey & (1)                     \\
Hartmann et al. (1991)    & 10  & Proper motion survey & (1),(2)    \\
Gomez et al. (1992)       & 2   & Proper motion survey & (1)        \\ 
Brice\~{n}o et al. (1993) & 12  & H$\alpha$ prism survey & (1)      \\ 
Leinert et al. (1993) 	  & 167 & nIR speckle interferometry        \\  
Ghez et al. (1993)        & 67  & Speckle imaging survey            \\   
Simon et al. (1995) 	  & 77  & IR lunar occultation survey       \\ \hline
\end{tabular}

\medskip
Notes: (1) as described in \S 2 of Gomez et al. (1993), (2) includes all the stars of their Table 1.
\end{minipage}
\end{table*}

For Chamaeleon I we use the positions of 137 optically visible T-Tauri stars compiled by Lawson, Feigelson and Huenemoerder (1996) and K-Band observations of binary systems by Ghez et al. (1997). For Taurus, positions of 216 optically visible pre-main-sequence stars have been compiled from the sources listed in Table \ref{tab:refer}; these are essentially the same data used by Gomez et al. (1993) but supplemented by the binary searches of Ghez et al. (1993), Leinert et al. (1993) and Simon et al. (1995).

\section {Surface-density of companions}
{\label{sec:surf_den}}

In order to study the distribution of young stars in Taurus, Gomez et al. (1993)  evaluated the two point angular correlation function. Larson (1995)  instead evaluated the mean surface-density of companions per star, $\bar{N}(\theta)$, as a function of angular separation, $\theta$; this measure avoids the need to define a finite background surface-density. Simon (1997) used both methods to study the distribution of stars in three different SFRs, and in each case calculated $\bar{N}(\theta)$ out to separations $\sim \frac{1}{10}$ of the overall angular extent of the region, where the two methods are in good agreement. Nakajima et al. (1998) also limited their fits to $\theta \leq \theta\mx$ with $\theta\mx$ less than the overall size of the SFR being analyzed; but they do not divulge how $\theta\mx$ was chosen, and our numerical experiments suggest that the fit can be quite sensitive to the choice of $\theta\mx$. 

\subsection{Data reduction}

To compute $\bar{N}(\theta)$, we take each star $n$ in turn, and divide the surrounding area of sky into a set of annuli, by drawing circles of radius $\theta_i$ centered on star $n$, with $\theta_{i} = 2\theta_{i-1}$ and $i \geq 1$. $\theta_0$ is chosen to be well below the smallest separation in the sample. Next we count the number of companion stars ${\cal N}_{n,i}$ in each annulus. Then,

\begin{equation}
\bar{N} \left( \bar{\theta}_i \right) \; = \; 
\frac{ \sum_{n} \left\{ {\cal N}_{n,i} \right\} }
{\pi {\cal N}\tl (\theta_{i}^{2} - \theta_{i-1}^{2})} \; . 
\end{equation}

\noindent where $\bar{\theta}_i \; = \; 
\left( \theta_i + \theta_{i-1} \right) / 2 \;$ and ${\cal N}\tl$ is the total number of stars in the cluster. 

Because of discreteness and small-number statistics, the results are quite sensitive to the choice of $\theta_0$. Therefore we repeat this procedure twice, each time increasing $\theta_0$ by $2^{1/3}$. This increases the number of plotted points threefold. It appears to resolve all the significant divergences between the plotted points and the fitting curves (see, for example, the dip around $\theta \sim 0.02^{\circ}$ in the plotted points for Taurus on Figure 1), and therefore there is nothing to be gained by increasing the number of points any further. We believe that it is important to resolve these divergences, at the very least because they provide a graphic measure of the limitations of the fitting curves. It may even be the case that these excursions constrain the previous evolutionary history of the SFR (cf. Nakajima et al. 1998, Bate et al. 1998).

\subsection{Corrections for incompleteness}

Since the stars in Chamaeleon I and Taurus have not all been surveyed for multiplicity, and since some of those which have been surveyed have been surveyed to different limits and/or by different techniques, we have attempted to correct for incompleteness. To do this we assume that multiplicity amongst young stars is scale-invariant, in the sense that a faint system is as likely to be a binary as a bright one. Hence, in Chamaeleon I we presume that the stars not surveyed for 
companions by Ghez et al. (1997) have -- on average -- the same multiplicity as those that were  surveyed. In Taurus we are guided by the multiplicity statistics for stars surveyed by Ghez et al. (1993), Leinert et al. (1993) and Simon et al. (1995). 
Systems are categorized according as they were surveyed in all three programmes, in two, in just one, or in none. They are then compensated for potential companions that would have been missed, on the assumption that the probability for companions in the ranges not surveyed is the same as for stars that were surveyed in these ranges. We emphasize that these corrections only change the coefficients of the fitting curves ($K\bn$ and $K\cl$ in Eqn. (\ref{simult})); they do not significantly affect the exponents ($\beta\bn$ and $\beta\cl$ in Eqn. (\ref{simult})).

Bate et al. (1998) discuss a Monte Carlo technique to correct for companions beyond the boundary of the survey region. However, this correction is only appropriate in cases where the cluster continues -- at a comparable surface-density -- outside the area of the survey. In the cases discussed here this is not the case, and so no such correction is attempted.

\subsection{Fitting procedure}

In order to fit $\bar{N}(\theta)$, we follow Simon (1997) in ignoring separations greater than one tenth of the angular extent of the region under consideration. We also ignore small separations where the counts are too low to afford reliable statistics. This leaves the ranges $0.000015^{\circ} < \theta < 0.15^{\circ}$ for Chamaeleon I; and $0.00006^{\circ} < \theta < 2.2^{\circ}$ for Taurus. The points outside these ranges are represented by open circles on Figure 1.

The procedure adopted by other workers has been to fit the observational data 
{\it piecewise} with two power laws. However, our experience has shown that the results are then rather sensitive to the breakpoint where the two power-laws meet, and there is no obvious strategy for choosing this breakpoint objectively. Moreover, we would -- in general -- expect there to be binaries with separations  greater than the elbow, and chance projections at separations less than the elbow.  Therefore our strategy has been to fit the observational data with two {\it simultaneous} power laws:

\begin{equation}
\label{simult}
\bar{N}(\theta) \; = \; K\bn \theta^{-\beta\bn} \, + \, K\cl \theta^{-\beta\cl} \; .
\end{equation}

The elbow $\theta\br$ is defined as the separation at which the two terms on the righthand side of Eqn. (\ref{simult}) are equal, i.e.

\begin{equation}
\theta\br = \mbox{exp} \left\{ 
\frac{\mbox{ln} \left[ K\bn \right] - \mbox{ln} \left[ K\cl \right]}
{\left[ \beta\bn - \beta\cl \right]} \right\} 
\end{equation}

In order to identify the best fit we evaluate for each trial fit the probability that the observational data could be generated from that fit by Gaussian statistics. The best fit is then the one with the highest probability, and the uncertainties embrace 1$\sigma$ departures from the best fit. The data that we fit are the corrected data, but it should be emphasized that we estimate the standard deviation for the contents of each annulus (and hence the differential probability) using the raw counts, and not the corrected counts.

\subsection{Results}

\begin{figure*}
         \psfig{figure=\figurepath/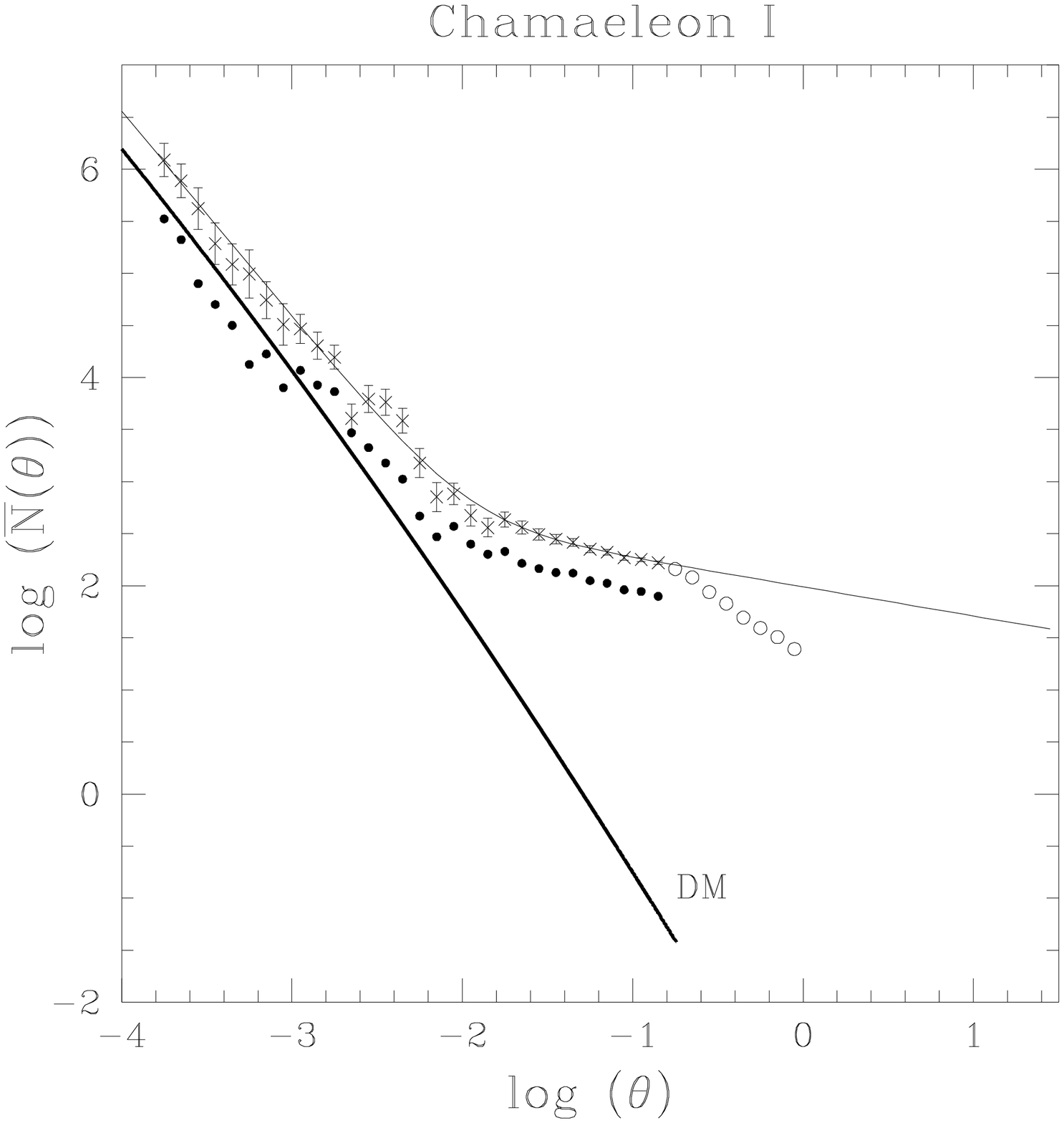,height=4.5in,width=4.5in}
         \psfig{figure=\figurepath/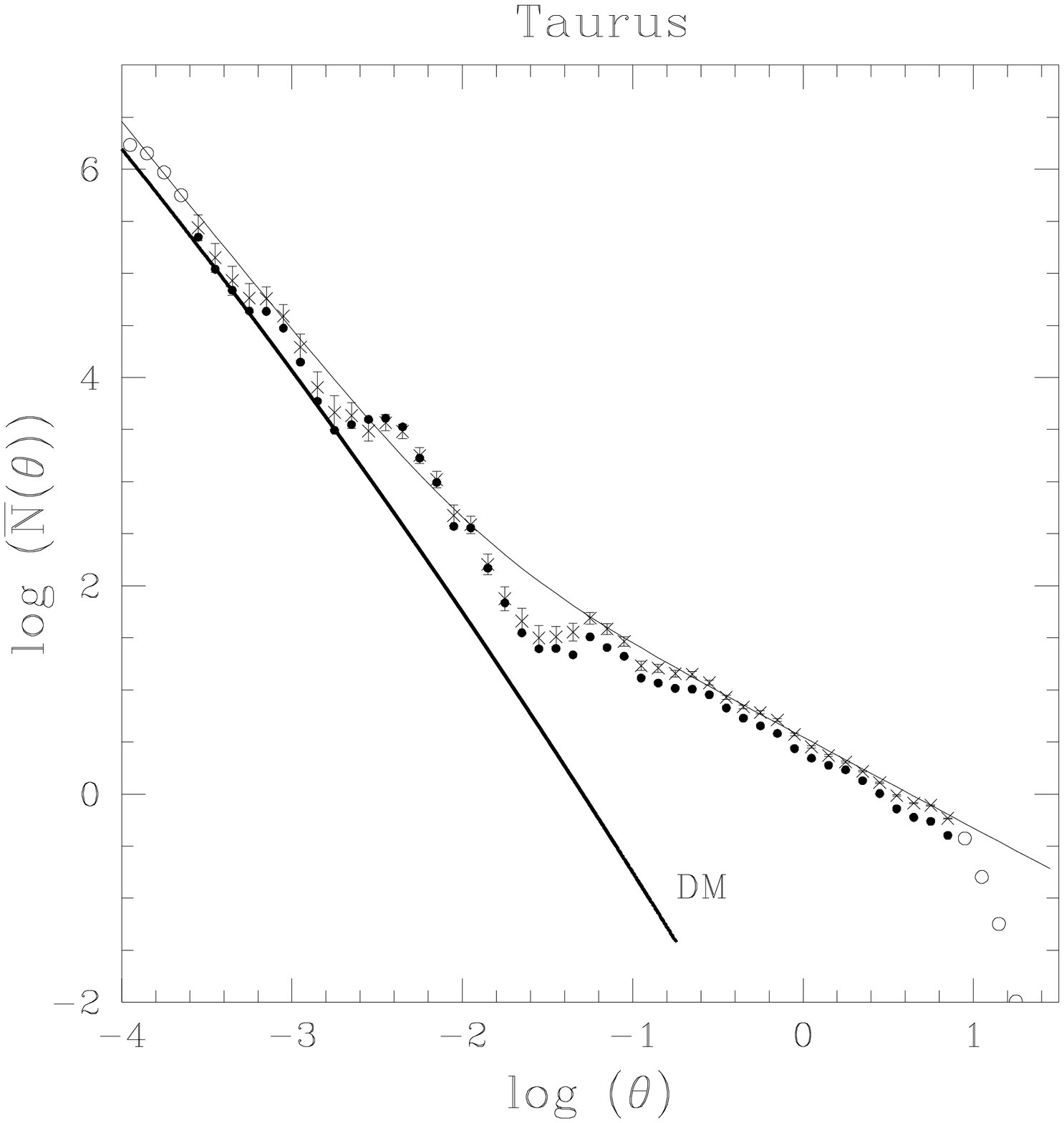,height=4.5in,width=4.5in} 
      \caption{Logarithmic plots of $\bar{N}(\theta)$ versus $\theta$ for Chamaeleon I (left) and Taurus (right). The crosses with 1$\sigma$ error bars are the corrected points which have been fitted with Eqn. (\ref{simult}) (thin solid curve). The open circles represent separations which are ignored in making the power-law fits (see discussion in Section 2.3). The filled circles are the uncorrected points. The thick solid lines marked DM show $\bar{N}(\theta)$ for companions to field stars, as obtained by Duquennoy \& Mayor (1991).}
      \label{fig:lar}
\end{figure*}

$\bar{N}(\theta)$ is plotted for Chamaeleon I and Taurus in Figure \ref{fig:lar}. The corrected points are the crosses with 1$\sigma$ error bars. For Chamaeleon I (lefthand panel of Figure \ref{fig:lar}), the best fit has exponents $\beta\bn = 1.97 \pm 0.07$ (in the binary/multiple regime at small separations) and $\beta\cl = 0.28 \pm 0.06$ (in the hierarchical clustering regime at large separations); the elbow is at $\theta\br \sim 0.011^{\circ} \pm 0.004 ^{\circ}$, corresponding to a projected separation of $s\br \sim$ 0.027 pc (5500 AU) at the distance of Chamaeleon I. For Taurus (righthand panel of Figure \ref{fig:lar}) the exponents are $\beta\bn = 2.02 \pm 0.04$ and $\beta\cl = 0.87 \pm 0.01$, with the elbow at $\theta\br \sim 0.013^{\circ} \pm 0.003 ^{\circ}$, corresponding to $s\br \sim$ 0.032 pc (6500 AU) at the distance of Taurus. Table \ref{table:comp} summarizes the values of $\beta\bn$, $\beta\cl$ and $s\br$ obtained by Larson (1995), Simon (1997), Nakajima et al. (1998), Bate et al. (1998), and this work.

We reiterate that the corrections we make for incompleteness have no significant effect on the exponents $\beta\bn$ and $\beta\cl$, or on the value of $\theta\br$. They tend to increase the coefficients $K\bn$ and $K\cl$ by 40-50$\%$. To demonstrate this we have plotted the uncorrected points as filled circles on Figure 1. Equally, if we count the multiple systems found in the surveys of Ghez et al. (1993), Leinert et al. (1993) and Simon et al. (1995) as single stars (or ``systems'') for the purpose of computing large separations, this too has no significant effect on $\beta\cl$. 

We note that in the binary regime ($\theta < \theta\br$) (a) the observational data make rather wide but apparently systematic excursions about the best fit line (up to 5$\sigma$); and (b) these same excursions are also present in the data presented by Larson (1995) and Simon (1997). They are more apparent in our plots simply because we have displayed more points. Such excursions might be evidence for evolutionary effects of the type described by Nakajima et al. (1998) and Bate et al. (1998), for example dispersal of unbound clusters, but equally they could have been present when the stars first formed, or they could be due to some unidentified selection effect. 

Duquennoy \& Mayor (1991) have measured the orbital elements of a large field sample of mature binary systems having F- and G-type primaries. They show that the distribution of periods is well fitted by a Gaussian,

\[
{\cal N}_{\mbox{log}_{\tiny 10}[P]} d\mbox{log}_{\tiny 10}[P] =
\]

\begin{equation}
\hspace{1.5cm} \frac{{\cal N}\tl}{(2 \pi)^{1/2} \sigma_{\mbox{\tiny o}}} \, 
\mbox{exp} \left\{ \frac{- \, \mbox{log}_{\tiny 10}^2 
\left[ P/P_{\mbox{\tiny o}} \right]}{2 \sigma_{\mbox{\tiny o}}^2} \right\} 
d\mbox{log}_{\tiny 10}[P] , 
\end{equation}

\noindent with $P_{\mbox{\tiny o}} \simeq$ 180 years and $\sigma_{\mbox{\tiny o}} \simeq 2.3$. If we assume that all these systems comprise a $0.8 M_\odot$ primary and a $0.2 M_\odot$ secondary in a circular orbit on the plane of the sky (i.e. we neglect the distributions of mass, eccentricity and orbital inclination), then the Duquennoy \& Mayor distribution corresponds to 

\begin{equation}
\label{DM}
\bar{N}_{\mbox{\tiny DM}}(\theta) = \frac{3 f_{\mbox{\tiny comp}}}
{2 (2 \pi)^{3/2} \mbox{log}_{\tiny e}[10] \sigma_{\mbox{\tiny o}}\theta^2}  
\, \mbox{exp} \left\{ \frac{- \, 9 \mbox{log}_{\tiny 10}^2 
\left[ \theta/\theta_{\mbox{\tiny o}} \right]}
{8 \sigma_{\mbox{\tiny o}}^2} \right\} ,
\end{equation}

\noindent where $f_{\mbox{\tiny comp}} \simeq 0.9$ is the mean number of companions per star, and $\theta_{\mbox{\tiny o}} \simeq 0.00006^{\mbox{\tiny o}}$. For comparison, this curve is plotted as a thick solid line on Figures 1a and 1b. The Gaussian part of Eqn. (\ref{DM}) has FWHM $\Delta \mbox{log}_{\tiny 10}[\theta] \simeq 3.6$, i.e. it is very flat. Therefore the dominant variation comes from the $\theta^{-2}$ term, and the Gaussian part simply superimposes a subtle negative curvature. 

It is well known that the Duquennoy \& Mayor field binary distribution predicts fewer wide binary systems ($\theta \ga 0.0001^{\circ}$) than are observed in Taurus, and Figure 1b confirms this. As pointed out by Padgett, Strom \& Ghez (1997), there are at least three evolutionary processes which might reduce the number of wide systems in a mature stellar population. (i) Tidal disruption should preferentially reduce the number of wide systems. (ii) Interactions with other stars may also harden some systems, moving them to smaller separations. (iii) Low-mass (substellar) companions may fade to invisibility.

Additionally, it may be that different SFRs generate different binary distributions, and then these are all convolved together to produce the Duquennoy \& Mayor field binary distribution. There is conflicting evidence on this issue. Padgett et al. (1997) find that in the range 138 to 1050 AU, the binary frequency in high-density SFRs in Orion is essentially the same as in the low-density SFRs studied by Reipurth \& Zinnecker (1993) -- and significantly higher than in the field. In contrast, Brandner \& K\"ohler (1998) show that in Sco-Cen there are significant variations in the distribution of binary separations in the range 15 to 450 AU, in the sense that sub-regions where there are more massive stars also tend to have closer binaries. It will be very important to establish how universal this trend is, for example by investigating whether the lack of massive stars in Chamaeleon I and Taurus is accompanied by a relative paucity of close binary systems. However, the small numbers of close systems in the data sets for Chamaeleon I and Taurus make it impossible to draw such a conclusion at this time.

\subsection{Projection effects}

We concur with Simon (1997) that the apparent shift in $s\br$ going from Taurus ($\sim 0.03$ to 0.06 pc) through Chamaeleon I ($\sim 0.03$ pc) and $\rho$-Ophiuchus ($\sim 0.02$ pc) to Orion-Trapezium ($\sim 0.002$ pc) is largely a projection effect. In other words, there appears to be an approximately universal distribution of separations in the binary regime, giving 

\begin{equation}
\bar{N}\bn \; \simeq \; K\bn \theta^{-\beta\bn} \; , 
\end{equation}

\noindent with $\beta\bn \, = \, 2.0 \pm 0.2 \;$ and $K\bn$ satisfying the normalization condition 

\begin{equation}
\int_{\theta = 0}^{\theta = \theta\br} \; \bar{N}\bn(\theta) \, 
2 \pi \theta d\theta \; \sim \; 1 \; ,
\end{equation}

\noindent i.e. an average of about one binary (or multiple) companion per star. Therefore in denser clusters the intercept between the binary regime and the clustering regime shifts to smaller projected separations.

However, the distribution of separations in the clustering regime appears to be much less universal. Specifically, $\beta\cl$, although always well defined, ranges from 0.0 (Simon's minimum for Orion-Trapezium) to 0.9 (our maximum for Taurus) -- see Table 2. Our experiments confirm the conclusions of Bate et al. (1998). The form of $\bar{N}(\theta)$ is very sensitive to the treatment of boundary effects, and to the sample used; this is well illustrated by the work of Nakajima et al. (1998). One possible reason for the very small $\beta\cl$ value we obtain for Chamaeleon I (as compared with Taurus) is that we have used an X-ray selected sample here and this may favour the diaspora of weak-line T Tauri stars (WTTS), as against the more centrally clustered classical T Tauri stars (CTTS). By contrast, in Taurus most of the stars we use were initially detected in emission-line surveys, and therefore we might expect CTTS to be better represented than WTTS, giving a large $\beta\cl$, as observed. It is also clear that the procedure of fitting the two power laws simultaneously can make a significant difference to the result. When we analyzed Simon's Taurus data by fitting two power laws separately, we obtained the same fitting parameters as he did.

\begin{table*}
\begin{minipage}{164mm}
\caption{Parameters of the power-law fits to $\bar{N}(\theta)$}
{\label{table:comp}}
\begin{tabular}{lcllllll}
SFR & Number & \multicolumn{2}{c}{binary regime} & break & \multicolumn{2}{c}{clustering regime} & Reference\footnote{Refs: L95: Larson (1995), S97: Simon (1997), BCM98: Bate, Clarke \& McCaughran (1998), NTHN98: Nakajima et al. (1998).} \\
 & of stars & & & & & &  \\
 & & log$_{10} \! \left[ \frac{K\bn}{\mbox{\tiny per deg}^2} \right]$
 & ~~~$\beta\bn$ & $s\br$/pc & log$_{10} \!
 \left[ \frac{K\cl}{\mbox{\tiny per deg}^2} \right]$ & ~~~~~ $\beta\cl$ &  \\ \hline
Trapezium$\;$ & 35 & -2.30 $\pm$ 2.52 & 2.13 $\pm$ 0.65 & 0.002 & ~4.65 $\pm$ 0.51 & 0.20 $\pm$ 0.21 &  S97 \\ 
Trapezium$\;$ & 744 &  & 2.00 & 0.003 & & 0.11 $\pm$ 0.02 & BCM98\footnote{Only results
using the McCaughrean et al. (1996) survey are quoted. If boundaries are taken into account, $\beta\cl$ becomes -0.02  $\pm$ 0.01.} \\
Lupus & 65 & -2.47 $\pm$ 0.46 & 2.2 $\; \pm$ 0.4 & 0.0053 & ~1.23 $\pm$ 0.23 & 0.82 $\pm$ 0.13 & NTHN98 \\
Cham. I & 103 & -2.11 $\pm$ 0.19 & 2.1 $\; \pm$ 0.2 & 0.013  & ~1.36 $\pm$ 0.03 & 0.57 $\pm$ 0.04 & NTHN98 \\ 
$\rho$ Oph. & 96 & -3.42 $\pm$ 0.32 & $2.5\; \pm 0.3$ & 0.026 & ~0.91 $\pm$ 0.03 & 0.36 $\pm$ 0.06 & NTHN98 \\
$\rho$ Oph. & 51 & -1.20 $\pm$ 0.41 & $1.88 \pm 0.10$ & 0.024 & ~1.44 $\pm$ 0.24 & 0.50 $\pm$ 0.19 &  S97 \\  
Taurus & ~\footnote{Larson used results from several different surveys, but
analyzed them separately, so there can be no unambiguous entry for his work in the second column of the table.} & -2.19 & 2.15 & 0.04 & ~0.53 & 0.62 & L95 \\  
Taurus & 80 & -1.65 $\pm$ 0.25 & 2.01 $\pm$ 0.06 & 0.06 & ~0.54 $\pm$ 0.15 & 0.64 $\pm$ 0.19 &  S97 \\ 
Cham. & 94 & -2.80 $\pm$ 0.16 & 2.4 $\; \pm$ 0.5 & 0.083  & -0.09 $\pm$ 0.02 & 0.55 $\pm$ 0.03 & NTHN98 \\
Orion OB & 361 & -1.66 $\pm$ 0.13 & 1.6 $\; \pm$ 0.4 & 0.13 & ~0.95 $\pm$ 0.005 & 0.15 $\pm$ 0.02 & NTHN98 \\
\hline
Cham. I & 137 & -1.33 $\pm$ 0.21 & 1.97 $\pm$ 0.07 & 0.027 & ~2.00 $\pm$ 0.04 & 0.28 $\pm$ 0.06 &  This work     \\ 
Taurus & 216 & -1.62 $\pm$ 0.10 & 2.02 $\pm$ 0.04 & 0.032 & ~0.54 $\pm$ 0.01 & 0.87 $\pm$ 0.01 &  This work \\ 
\hline 
\end{tabular}
\end{minipage}
\end{table*}

\subsection{Box dimension}

We have also calculated the box-dimensions of these two SFRs. To do this we place a uniform Cartesian grid over the star field and count the number of grid-cells ${\cal N}\oc$ which contain stars. This is repeated for several different values of the grid-spacing $\Delta \theta$.

Strictly, the box-dimension is given by

\begin{equation}
D\bx \; = \; \lim_{\Delta \theta \rightarrow 0} 
\left\{ \frac{dln[{\cal N}\oc]}{dln[1/\Delta \theta]} \right\} \; . 
\end{equation}

\noindent However, since the clustering with which we are concerned has a finite dynamic range, we have adopted the following procedure.

First we determine the range of $\Delta \theta$ values to be considered. To do this we construct a square box of side $\Delta \theta\mx$ which just contains the cluster under investigation. If the cluster contains ${\cal N}\tl$ stars, we replace them with ${\cal N}\tl$ points distributed {\it randomly} within the boundaries of this box. Next we find the smallest grid-spacing $\Delta \theta\mn$ for which the box-dimension of the randomly distributed points equals 2. The range of $\Delta \theta$ values is then taken to be $\Delta \theta\mn \, \leq \, \Delta \theta \, \leq \, \Delta \theta\mx$.

Finally we return to the star cluster and evaluate the slope of the plot of log$[{\cal N}\oc]$ against log$[\Delta \theta]$ for $\Delta \theta$ values in this range:

\begin{equation}
D\bx \; = \; \left. \left< \frac{dln[{\cal N}\oc]}{dln[1/\Delta \theta]} \right> \right|_{\; \Delta \theta\mn \leq \Delta \theta \leq \Delta \theta\mx} \; .
\end{equation}

\noindent For Chamaeleon I this gives $D\bx \, = \, 1.51 \pm 0.12$ in the range $0.29^\circ \leq \Delta \theta \leq 2.0^\circ$; and for Taurus, $D\bx \, = \, 1.39 \pm 0.01$ in the range $3^\circ \leq \Delta \theta \leq 20^\circ$.

Given the small dynamic range of $\Delta \theta$, the significance of $D\bx$ is simply as a descriptor for the structure on the largest scales. It has no significance unless quoted along with the dynamic range of $\Delta \theta$. In particular it does not constitute admissible evidence of fractality.

\section {Surface-density mapping algorithms}
{\label{sec:clus}}

Gomez et al. (1993) have generated surface-density maps of the young 
stars in Taurus by applying a kernel smoothing technique. With this technique they demonstrate the existence of several sub-groups of Pre-Main--Sequence stars  within the Taurus cluster. Such maps are useful for comparing the distribution of young stars with the distribution of dense gas from which stars form, and also for comparing with maps generated by numerical simulations of star-formation. However, any such map depends critically on the choice of smoothing length. It is unclear how Gomez et al. made this choice. We present here two algorithms for determining the smoothing length objectively. Algorithm I ({\it SCATTER}) uses a universal smoothing length. Algorithm II ({\it GATHER}) uses a local smoothing length. We stress that although the final maps are evidently similar morphologically, this does not mean that the algorithm used to generate a map is unimportant. The information extracted from the raw star field is substantially different. The principal differences lie in the contour levels, which have a much larger dynamic range for Algorithm II ({\it GATHER}).

In both cases a grid is drawn over the star field, and the 
surface-density of stars is calculated at each 
grid point. Standard software is then used to plot 
contours of constant surface-density. Provided the grid 
is sufficiently fine, the precise grid spacing has no 
significant influence on the final map.

\section{Algorithm I: {\it SCATTER}}

Given ${\cal N}\tl$ stars with positions ($x_{n},y_{n}$) $\; (n = 1$ to ${\cal N}\tl$), Algorithm I evaluates the arithmetic mean ($\bar{s}$) of their angular separations:

\begin{eqnarray}
\label{equa:sbar}
\bar{s} & = & \frac{2}{{\cal N}\tl({\cal N}\tl-1)} 
\sum_{n=1}^{n=({\cal N}\tl-1)} \sum_{n'=(n+1)}^{n'={\cal N}\tl} \nonumber \\ 
 & & \left\{ \left[ (x_n - x_{n'})^2 + (y_n - y_{n'})^2 \right]^ \frac{1}{2} \right\} \; .
\end{eqnarray}

\noindent The universal smoothing length $h$ is then given by

\begin{equation}
h = \left[ \frac{2}{{\cal N}\tl} \right]^{1/2} \, \bar{s} \; .
\end{equation}

Each star $n$ is considered in turn and distributed to the surrounding grid points $(x,y)$ using a normalized linear smoothing kernel with compact support, 

\begin{equation}
W(r) \; = \; \left\{ \begin{array}{ll}
3 ( h - r ) / \pi h^3 \; , \hspace{1cm} & r \, < \, h \; ; \\
0 & r \, > \, h \; ,
\end{array} \right.
\end{equation}

\noindent where $r^2 \, = \, (x-x_n)^2 + (y-y_n)^2$.

The contributions from individual stars are summed at each grid point to obtain a surface-density array, and then contour diagrams are plotted. The minimum contour is drawn at 

\begin{equation}
\label{eq:sig}
N_{\mbox{\tiny min}} = \frac{{\cal N}\tl}{2 \bar{s}^{2}} \; ;
\end{equation}

\noindent so that isolated single stars are not contoured, i.e. $N_{\mbox{\tiny min}} > W(0)$. 

Since the kernel is normalized, the final surface-density map is also normalized, in the sense that

\begin{equation}
\label{norm}
\int_{\mbox{\tiny the sky}} \; N(x,y) dx dy \; = \; {\cal N}\tl \; .
\end{equation}

\vspace{0.2cm}

However, by using a universal smoothing-length $h$, we are inevitably capturing only structures on scales of order $h$. Structures on scales much smaller than $h$ are smoothed out, because the constituent stars are closer together than their individual smoothing-lengths. Structures on scales much greater than $h$ are not contoured because the kernels of their constituent stars do not overlap. If clustering is hierarchical (e.g. Larson 1995), the limited dynamical range of the {\it SCATTER} algorithm will suppress this fact.

Nonetheless, the resulting contour maps appear to reproduce the features which the human eye sees. This is demonstrated in Figures 2 and 3. The raw star fields are shown in Figure 2, with Chamaeleon I at the top, and Taurus at the bottom. The contour maps obtained with the {\it SCATTER} algorithm are shown in Figure 3. Note that the linear scale of the Chamaeleon I map ($\sim \, 1^{\circ}$) is about ten times smaller than the linear scale of the Taurus map ($\sim \, 10^{\circ}$).

\section{Algorithm II: {\it GATHER}}

In Algorithm II we compute for each grid point, the distance 
$d_k$ to the $k^{\mbox{\tiny th}}$ nearest-neighbour star, for 
$k \, \leq \, k_{\mbox{\tiny max}}$. The surface-density 
$N_{\mbox{\tiny g}}$ at that grid point is then calculated 
by gathering contributions from these stars according to

\[
N_{\mbox{\tiny g}} \; = \; \left( \sum_{k = 1}^{k = k_{\mbox{\tiny max}}} 
\, \left\{ \frac{w_k \; \pi \; d_k^2}{k- \frac{1}{2}} \right\} \right)^{-1} \; .
\]

\noindent Thus the effective smoothing length is an average of the distances $d_k$ to the nearest $k_{\mbox{\tiny max}}$ stars. The best results are obtained with a weighted sum of the squared distances.

By invoking an adaptive local smoothing length, this algorithm ensures that the computed surface-density is 
finite everywhere -- albeit very low in places where the stars are 
few and far between. This extends the dynamic range of the algorithm, and thereby increases its chance of capturing any hierarchical structure in the underlying star field. 

Algorithm II is objective provided that the choice of 
$k_{\mbox{\tiny max}}$ and the scheme for computing the weighting 
factors $w_k$ are objective. For simplicity we use:

\begin{equation}
w_k \; \propto \; \mbox{exp} \left[ - |k-4| \right] \; ,
\end{equation}

\noindent with

\vspace{0.1cm}

\begin{equation}
\sum_{k = 1}^{k = k_{\mbox{\tiny max}}} \, 
\left\{ w_k \right\} \; = \; 1 \; ,
\end{equation}

\vspace{0.1cm}

\noindent and $k_{\mbox{\tiny max}} \, = \, 7$. Two considerations inform this choice. First, 
very few grid points will have more than 7 {\it natural-neighbour} 
stars\footnote{If the grid point and its nearby stars are treated as 
nodes, then the natural-neighbour stars of the grid point are 
those stars whose Voronoi cells share a boundary with the Voronoi 
cell of the grid point, {\it e.g.} Braun \& Sambridge (1995)}. 
Second, we do not wish to weight heavily the closest neighbouring stars, otherwise we will over-emphasize binary systems.

However, the resulting surface-density map is only normalized, in the sense of Eqn. (\ref{norm}), if the surface-density is truncated at a finite value. For Chamaeleon I this truncation value is $N_{\mbox{\tiny min}} = 47.7$ stars per square degree, and for Taurus it is $N_{\mbox{\tiny min}} = 4.06$ stars per square degree. In other words, at all grid points where $N_g$ falls below the truncation value, we set $N_g$ to zero.

Again the resulting contour maps appear to reproduce the features which the human eye sees. This is demonstrated in Figure 4. Inspection of the limiting contour values listed in Table 3 shows that the {\it GATHER} algorithm does indeed have a significantly larger dynamic range than the {\it SCATTER} algorithm. For this reason {\it GATHER} is our prefered algorithm.

\begin{figure*}
         \psfig{figure=\figurepath/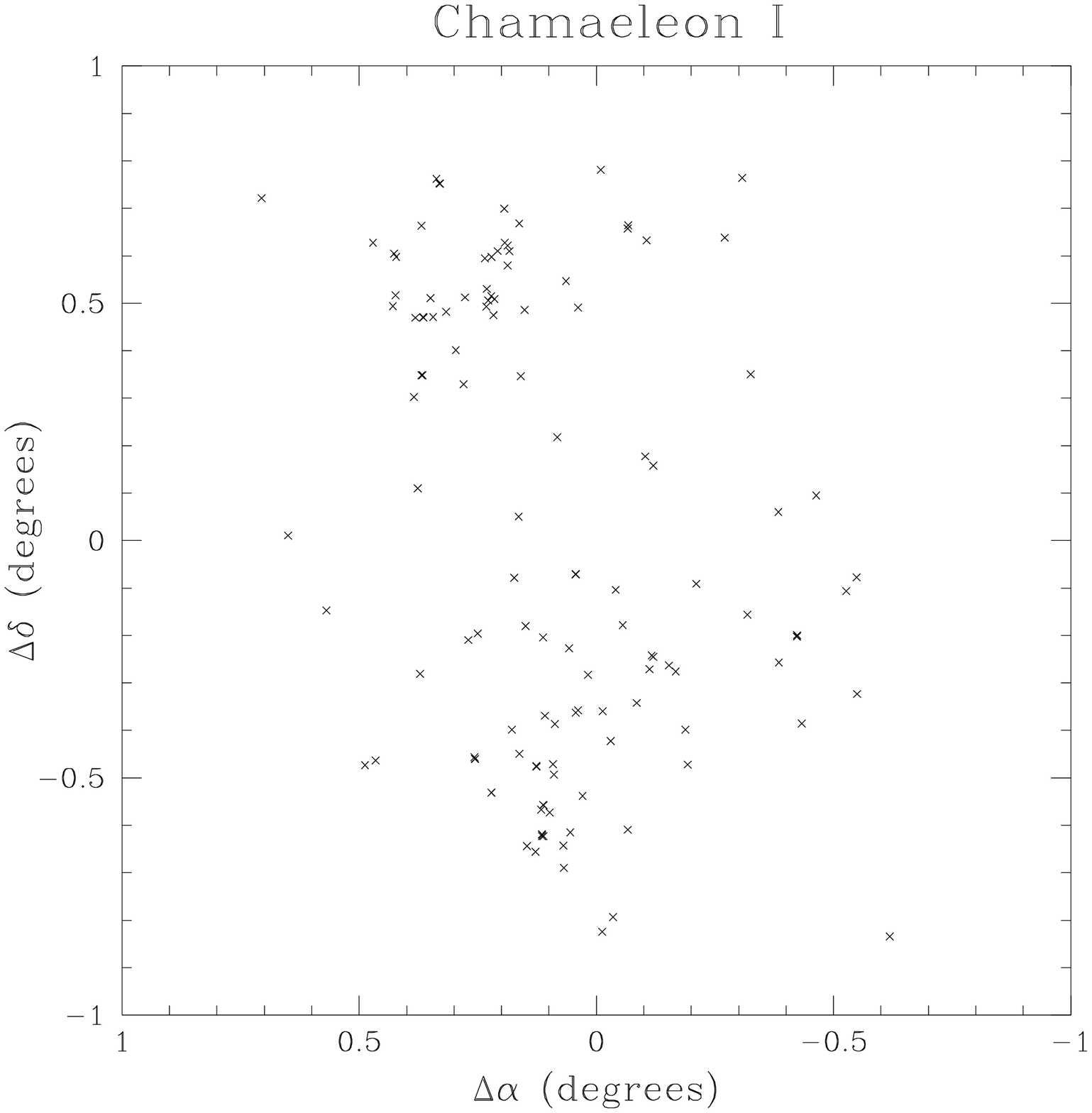,height=4.5in,width=4.5in} 
         \psfig{figure=\figurepath/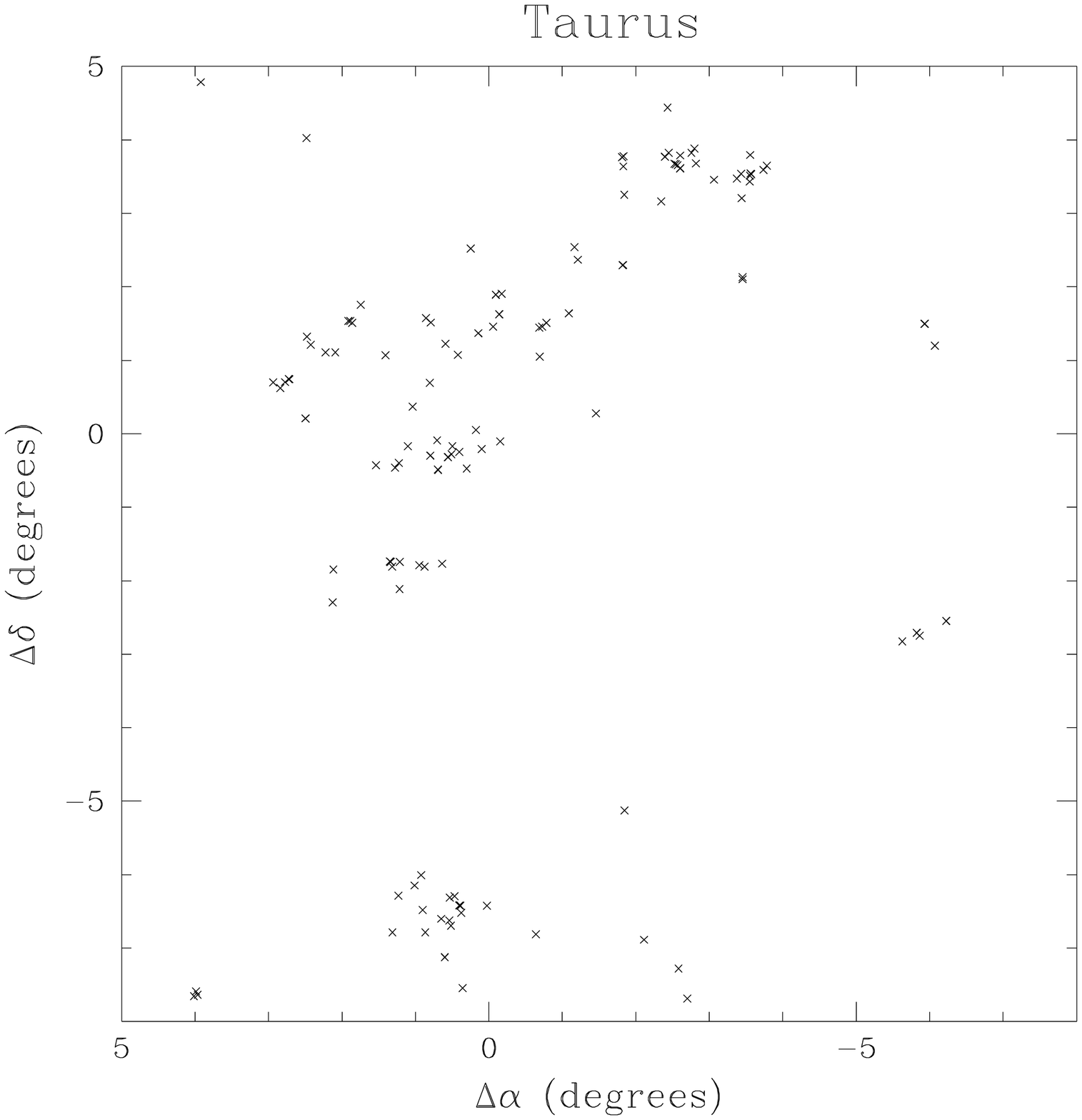,height=4.5in,width=4.5in} 
      \caption{The star positions for Chamaeleon I (top) and Taurus (bottom). The plots are centred at $\alpha$ = 11 hrs 7 mins 45.6 secs and $\delta$ = - 77$^{o}$ 5$'$ 24$''$ for Chamaeleon I and at $\alpha$ = 4 hrs 27 mins 2.4 secs and $\delta$ = 24$^{o}$ 33$'$ for Taurus.}
      \label{fig:stars}
\end{figure*}

\begin{figure*}
         \psfig{figure=\figurepath/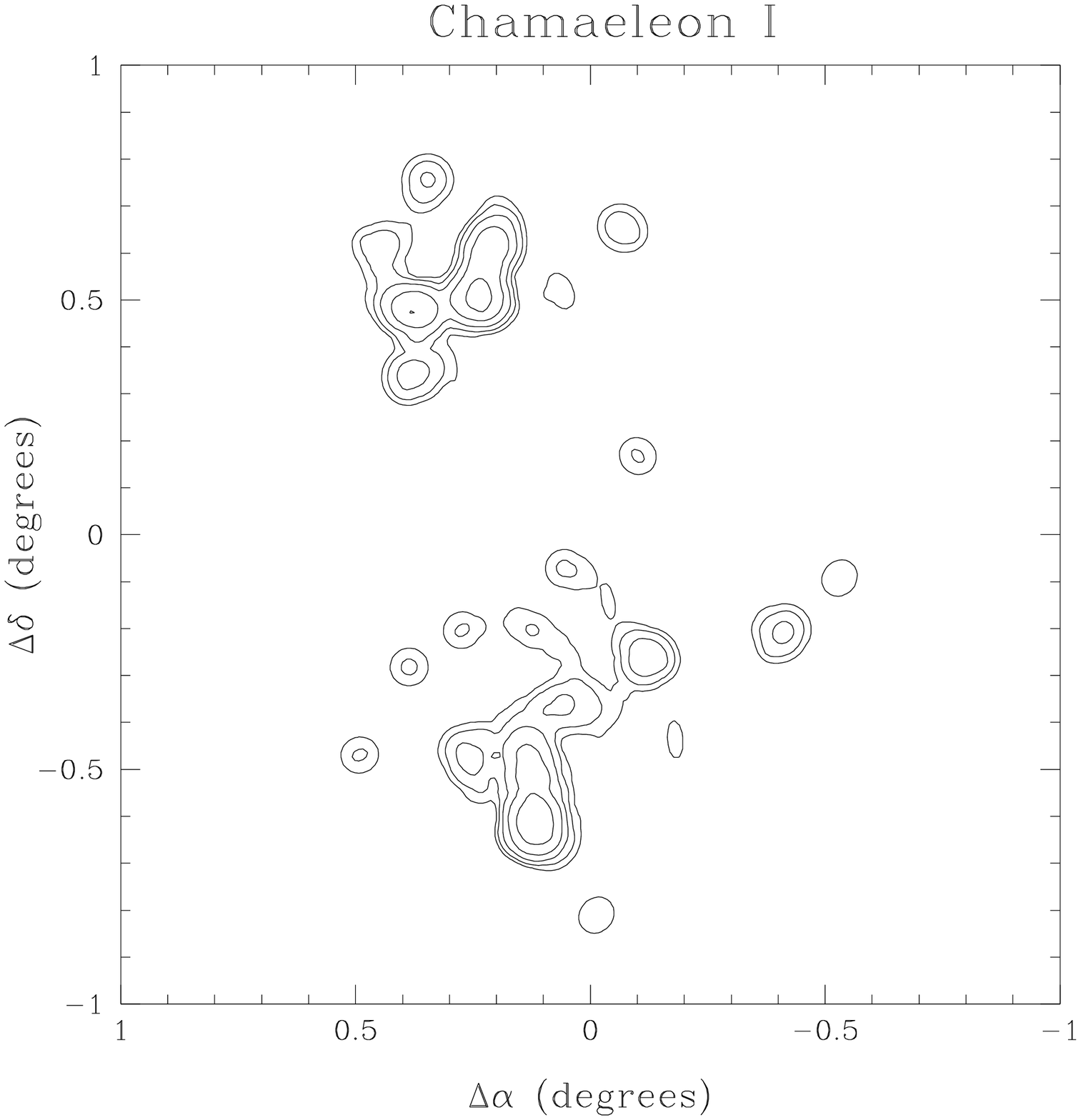,height=4.5in,width=4.5in} 
         \psfig{figure=\figurepath/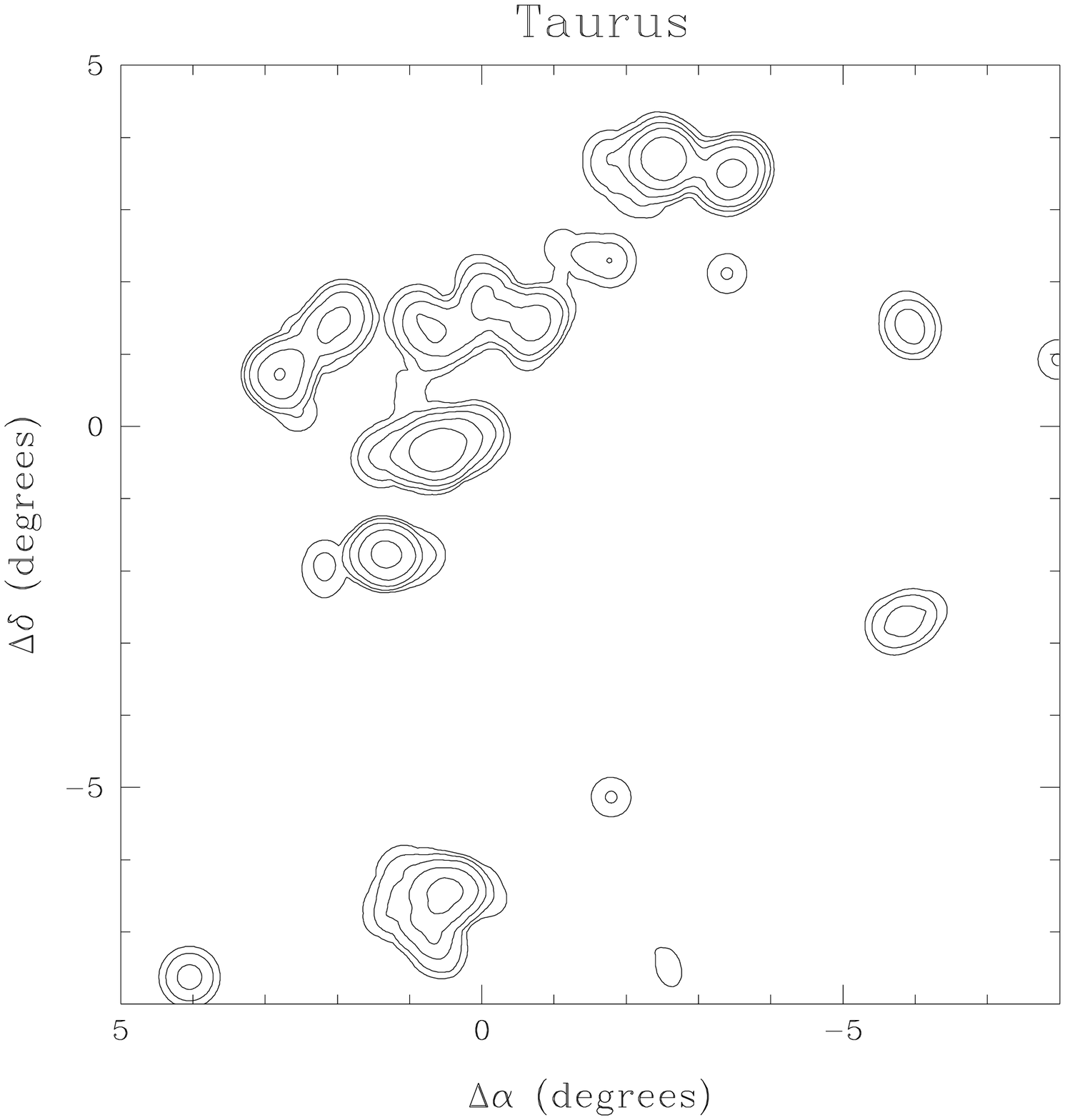,height=4.5in,width=4.5in} 
      \caption{The surface-density plots generated by {\it SCATTER}, for Chamaeleon I (top) and Taurus (bottom). The plots are centred at $\alpha$ = 11 hrs 7 mins 45.6 secs and $\delta$ = - 77$^{o}$ 5$'$ 24$''$ for Chamaeleon I and at $\alpha$ = 4 hrs 27 mins 2.4 secs and $\delta$ = 24$^{o}$ 33$'$ for Taurus.}
      \label{fig:cont}
\end{figure*}

\begin{figure*}
         \psfig{figure=\figurepath/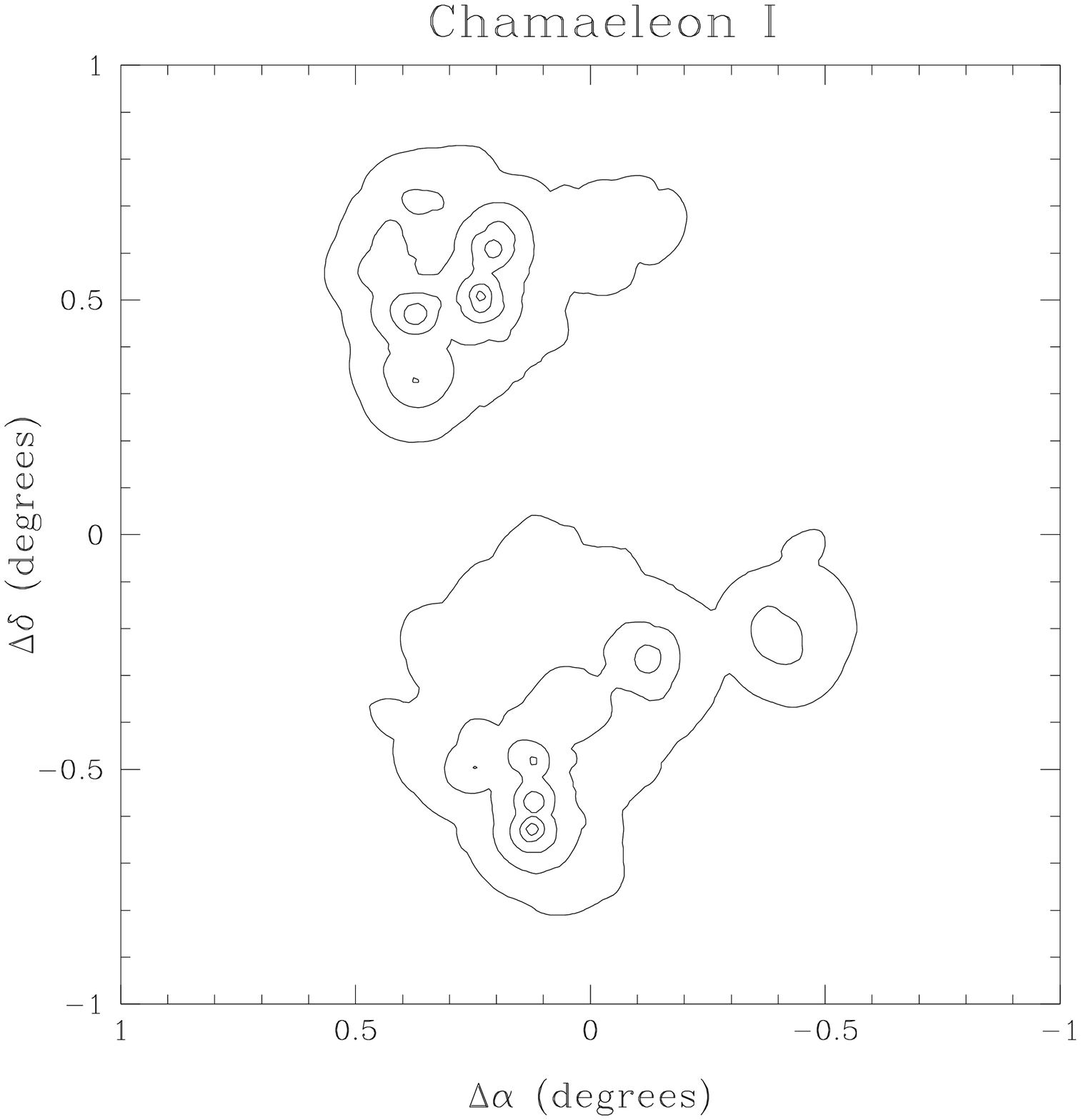,height=4.5in,width=4.5in} 
         \psfig{figure=\figurepath/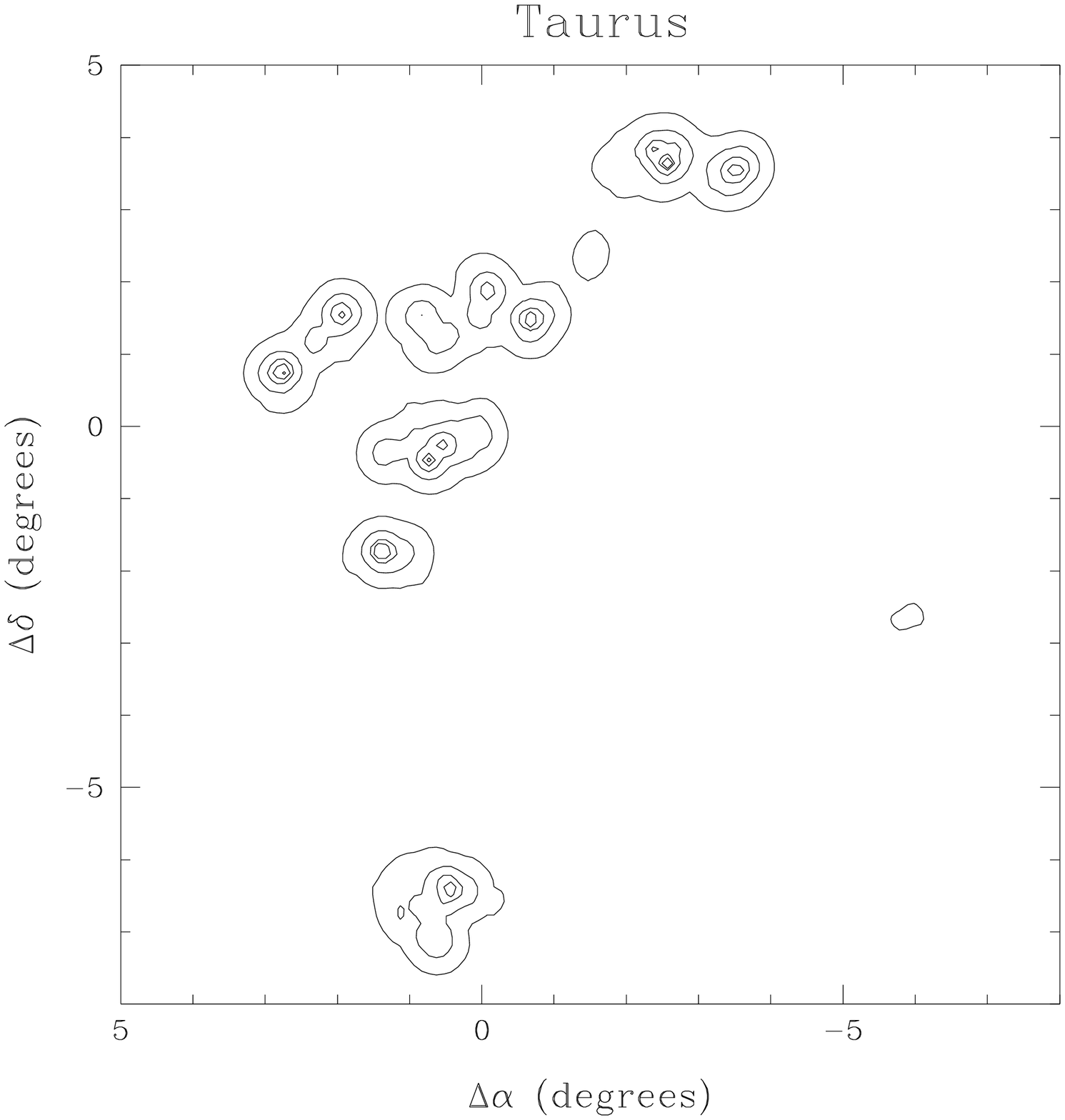,height=4.5in,width=4.5in} 
      \caption{The surface-density plots generated by {\it GATHER}, for Chamaeleon I (top) and Taurus (bottom). The plots are centred at $\alpha$ = 11 hrs 7 mins 45.6 secs and $\delta$ = - 77$^{o}$ 5$'$ 24$''$ for Chamaeleon I and at $\alpha$ = 4 hrs 27 mins 2.4 secs and $\delta$ = 24$^{o}$ 33$'$ for Taurus.}
      \label{fig:gath}
\end{figure*}

\section{Results}

Contour maps obtained by the two algorithms described above are shown in Figures 3 and 4. On all four maps, the contours are equally spaced in logarithm between $N_{\mbox{\tiny min}}$ and $N_{\mbox{\tiny peak}}$, but since $N_{\mbox{\tiny peak}}$ corresponds to the highest single pixel, the maximum contour drawn falls below $N_{\mbox{\tiny peak}}$ at $N_{\mbox{\tiny max}}$. On all maps, only 5 contours are drawn, to avoid confusion.

We see that Chamaeleon I divides into two main sub-clusters, having $>$ 100 stars per square degree, and linear extent $\sim$ 0.4$^{\circ}$ or $\sim$ 1 pc. Within these are three more compact sub--sub-clusters having $>$ 500 stars per square degree, and linear extent $\sim$ 0.1$^{\circ}$ or $\sim$ 0.25 pc. Taurus contains many more sub-clusters. Gomez et al. (1993) identify six sub-clusters, but of course the precise number depends on how a sub-cluster is defined. If we require a sub-cluster to be an isolated peak of at least one order of magnitude in surface density, we find 7 sub-clusters in the {\it GATHER} map -- which is the more discriminating map. The sub-clusters in Taurus are much less dense (typically $\sim$ 10 stars per square degree) than those in Chamaeleon I.

In particular we note that the two Chamaeleon I sub-clusters are similar in linear size and separation to the two sub-clusters in the north-west corner of Taurus (Groups I and II of Gomez et al. 1993, at $\alpha \simeq $ 4hrs 15mins and $\delta \simeq 28^\circ$). However, the two sub-clusters in Chamaeleon I contain about five times as many stars as Groups I and II in Taurus.

In fact the sub-clusters in Chamaeleon I are probably bound. Each sub-cluster contains about 50 stars, so its total mass is $\sim 10^{35}$ g, and each sub-cluster has a radius $\sim 0.5$ pc. Therefore they are bound provided their intrinsic velocity dispersion is $\Delta v \la 1$ km s$^{-1}$.

\begin{table*}
\begin{minipage}{130mm}
\caption{Contour limits and peak values for the maps in Fig. 2}
{\label{table:cont}}
\begin{tabular}{@{}lcccccccccc@{}}
Algorithm & $\;\;\;\;\;\;\;$ & \multicolumn{4}{c}{Chamaeleon I} & $\;\;\;\;\;\;\;$ & \multicolumn{4}{c}{Taurus} \\
 & & $\bar{N}_{\mbox{\tiny min}}$ & \mbox{range} & $\bar{N}_{\mbox{\tiny max}}$
 & $\bar{N}\pk$ & & $\bar{N}_{\mbox{\tiny min}}$ & \mbox{range} & $\bar{N}_{\mbox{\tiny max}}$ & $\bar{N}\pk$ \\ \hline
{\it SCATTER} & & 140 & $\times 8.2$ & 751 & 1142 & & 3 & $\times 11.7$ & 21 & 35\\
{\it GATHER} & & 47.7 & $\times 287$ & 4413 & 13690 & & 4.06 & $\times 256$ & 342 & 1038 \\  \hline
\end{tabular}

\medskip
On all plots there are 5 contours, and the contours are equally spaced logarithmically between $\bar{N}_{\mbox{\tiny min}}$ and $\bar{N}\pk$.
\end{minipage}
\end{table*}

\section{Conclusions}

We have calculated the surface-density of companions per star as a function of angular separation for Chamaeleon I and Taurus. We have used a fitting procedure which avoids the subjectivity of locating the elbow by eye. For Chamaeleon I , we obtain $\bar{N}(\theta) \propto \theta^{-1.97}$ for $\theta \ll 0.013^{\circ}$ and $\bar{N}(\theta) \propto \theta^{-0.28}$ for $\theta \gg 0.013^{\circ}$. For Taurus, we obtain $\bar{N}(\theta) \propto \theta^{-2.02}$ for $\theta \ll 0.011^{\circ}$ and $\bar{N}(\theta) \propto \theta^{-0.87}$ for $\theta \gg 0.011^{\circ}$.

In combination with the results of Larson (1995), Simon (1997) and Nakajima et al. (1998), our results suggest that universal processes are at work determining the multiplicity of newly formed stars, with $\beta\bn \simeq 2.0 \pm 0.1$, and $K\bn \simeq 3. \pm 1. \times 10^{-2}$ stars per square degree. One possibility is that the statistics of binary and multiple systems are determined by dynamical fragmentation occuring in gas which has suddenly become dense enough to be cooled efficiently by dust (Whitworth, Boffin \& Francis 1998).

In contrast, the large-scale clustering of young stars shows quite a large variation from one star-formation region to another. In general the power-law fit to $\bar{N}(\theta)$ in the clustering regime is rather tight, and hence the values of $K\cl$ and $\beta\cl$ obtained are well determined, although they do depend quite sensitively both on the sample used, and on the analysis technique used. The differences between star-formation regions are not solely due to different overall surface densities. There is no compelling evidence for a universal clustering process at work. 

We have also calculated the box-dimensions for the two star clusters: $1.51 \pm 0.12$ in the range $0.29^\circ \leq \Delta \theta \leq 2.0^\circ$ for Chamaeleon I, and $1.39 \pm 0.01$ in the range $3^\circ \leq \Delta \theta \leq 20^\circ$ for Taurus. We emphasize that these are only descriptors of the clustering and not admissible evidence for fractality.

We have presented two objective algorithms for constructing surface-density 
maps of star clusters. The {\it SCATTER} algorithm has the advantage that it is implicitly normalized, but it has limited dynamic range and therefore suppresses hierarchical clustering. The {\it GATHER} algorithm has the advantage of increased dynamic range, but it has to be normalized explicitly. Both algorithms capture the structures seen by the human eye, and both avoid contouring individual stars 

The surface-density maps show that the sub-clusters in Chamaeleon I are very similar in linear size ($\sim$ 1 pc) and separation ($\sim$ 2 pc) to those in Taurus, but apparently much denser; if we assume that the sub-clusters in both SFRs are approximately spherical, then those in Chamaeleon are denser than those in Taurus by a factor $\ga$ 5. There appears to be a characteristic length-scale for cluster formation which is roughly independent of cluster mass. For instance, Lada et al. (1991) find that the embedded clusters in L1630 also have sizes and separations of this order.

One of the interesting problems which Bate et al. (1998) identify is that a given $\beta\cl$ could derive from a fractal distribution, but it could equally derive from single-level clustering. The resolution of this dichotomy may lie in the analysis of surface-density maps of the type we have generated here. For instance, in principle fractality should be revealed by a plot of perimeter {\it versus} area for the iso--surface-density contours. However, in practice, we have found that in Taurus and Chamaeleon I there are insufficient contours to obtain a reliable dimension. And in reality, the notion of fractal dimension may not be very meaningful for a property (the surface density) with such a small dynamic range, or a star field with so few stars ($\sim$ 100).

Another problem which Bate et al. (1998) identify is that $\bar{N}(\theta)$ does not give any indication of the density profiles of individual sub-clusters. Here surface-density contours can clearly help, since even small-$N$ clusters are resolved in unconfused regions like Taurus. For an isolated cluster defined by contours at surface-densities $N_k$ (with $N_{k+1} > N_k$) and enclosing areas 
$A_k$, we have

\begin{equation}
\left. - \; \frac{dln[N]}{dln[R]} \right|_{\mbox{ \tiny central}} = 
\lim_{k \rightarrow k\mx}  \left\{ \frac{A_k^{1/2} 
\left( N_k - N_{k-1} \right) }
{N_k \left( A_{k-1}^{1/2} - A_k^{1/2} \right) } \right\} 
\end{equation}

\noindent and

\begin{equation}
\left. - \; \frac{dln[n]}{dln[R]} \right|_{\mbox{ \tiny central}} \; = \; 
\left. - \; \frac{dln[N]}{dln[R]} \right|_{\mbox{ \tiny central}}  + \; 1 \; , 
\end{equation}

\noindent where $k\mx$ is the central contour in the cluster, $N$ is the surface-density, and $n$ is the volume-density.

\section*{Acknowledgements}

PPG gratefully acknowledges the receipt of a PPARC studentship, and HMJB the support of PPARC grant GR/K 94157. We thank the referee for suggesting that we make an explicit comparison between our distributions and the results of Duquennoy \& Mayor for field stars.

\end{document}